\documentclass[11pt]{article}

\usepackage[final]{acl}

\usepackage{times}
\usepackage{latexsym}

\usepackage[T1]{fontenc}

\usepackage[utf8]{inputenc}

\usepackage{microtype}

\usepackage{inconsolata}

\usepackage{graphicx,amsmath,multirow,makecell,subcaption,booktabs,tabularx,threeparttable}
\usepackage{hyperref}
\hypersetup{
    colorlinks=true,            
    linkcolor=blue,             
    filecolor=magenta,          
    urlcolor=cyan,              
    pdftitle={Overleaf Example},
    pdfpagemode=FullScreen,
    }

\usepackage{enumitem}
\usepackage[normalem]{ulem}

\def\wen#1{}
\title{UniVocal: Unified Speech-Singing Code-Switching Synthesis}

\author{
  Yufei Shi$^1$ \quad Qian Chen$^1$ \quad Wen Wang$^1$ \quad Xiangang Li$^1$ \quad Zhen-Hua Ling$^2$ \quad Yang Ai$^2$ \\
  $^1$Tongyi Fun Team, Alibaba Group \\
  $^2$Independent Researcher \\
  \texttt{zkddsr2023@mail.ustc.edu.cn, \{lukechan1231, wwang.969803\}@gmail.com}}

\begin{document}
\maketitle

\begin{abstract}
\wen{Please check my suggestion on the story sent in the dingtalk group.}
We propose \textbf{UniVocal}, a unified framework that implicitly infers vocal modes from text context to pioneer \textbf{Speech-Singing Code-Switching (SCS) Synthesis}—a task where transitions are autonomously driven by textual semantics, akin to seamless human language blending. Unlike single-mode generation or systems relying on switching-control tags, our proposed \textit{UniVocal} implicitly infers vocal modes solely from text context. To achieve this, we employ a data-efficient two-stage curriculum learning strategy that progressively trains a competitive TTS system to acquire the desired SCS capability. Addressing data scarcity, we introduce a scalable pipeline to synthesize diverse code-switching data that is both semantically and acoustically natural, alongside a new multi-scenario benchmark, \textbf{SCSBench}. To address limitations of semantic tokenizers in capturing acoustic details, we also introduce refined cent token and Chain-of-Thought (CoT) generation for planning prosody before content generation, effectively enhancing empathetic speech generation and singing melody.  Experimental results demonstrate that \textit{UniVocal} achieves state-of-the-art performance on \textbf{SCSBench} while maintaining competitive performance on regular speech and singing tasks. Audio samples are available at \url{https://project-univocal-demo.github.io/demo/}. The code and dataset are released at \url{https://github.com/FunAudioLLM/FunResearch/tree/main/UniVocal}.

\end{abstract}

\section{Introduction}

Recent advances in neural architectures and large-scale training data have enabled significant progress in speech and singing synthesis \citep{du2024cosyvoice2,zhang2025tcsinger2,ji2024wavchat}. However, a natural vocal behavior remains largely unexplored: seamless code-switching between speech and singing within a single utterance. In daily communication, humans instinctively blend these vocal modes based on semantic context, such as casually humming a melody during conversation, incorporating melodic fragments into storytelling, or using song to aid memory in educational settings. Yet existing audio generation systems cannot automatically determine when to switch between modes based solely on text content. We define this capability as \textbf{Speech-Singing Code-Switching (SCS) Synthesis}, the task of generating vocal streams where speech and singing automatically switch based on textual semantics.

\begin{figure*}[htbp]
    \centering
    \includegraphics[scale=0.3]{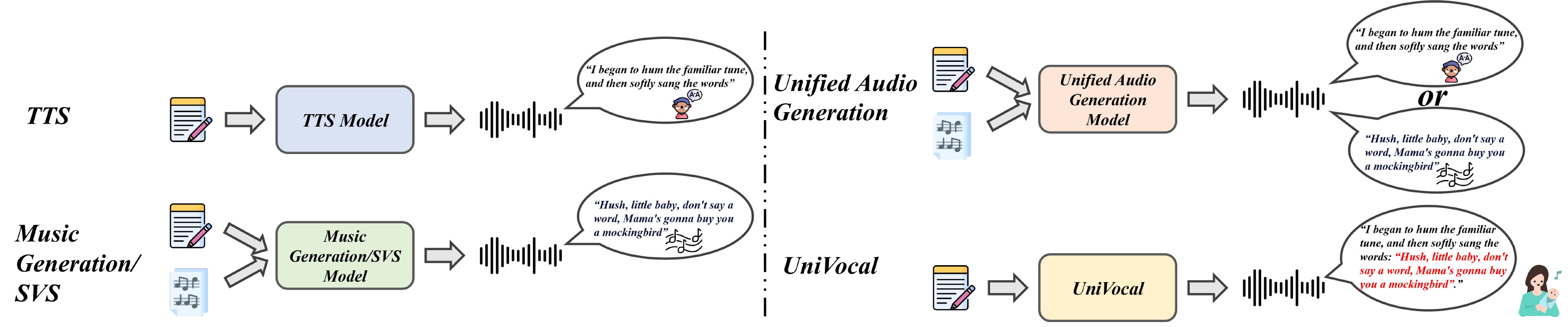}
    \caption{Common audio generation tasks, categorized into specialized tasks on the left and unified tasks on the right. Notably, in addition to its capabilities for regular speech and singing generation, \textit{UniVocal} can also produce vocal streams where speech and singing naturally code-switch.}
    \label{fig:audio_generation}
\vspace{-10pt}
\end{figure*}

Existing audio generation approaches, as illustrated in Figure \ref{fig:audio_generation}, are highly specialized and not suitable for the hybrid nature of SCS. Text-to-speech (TTS) systems \citep{du2024cosyvoice2,zhou2025indextts2} are limited to spoken prosody, lacking melodic expression. Similarly, music generation \citep{liu2025songgen,lei2025levo} and singing voice synthesis (SVS) \citep{zhang2024tcsinger,zhang2025tcsinger2} prioritize following musical rules or provided scores over linguistic content, restricting them to the singing mode. Current unified audio generation frameworks \citep{lei2023unisyn,yang2023uniaudio,zhang2025vevo2} generate only speech or singing based on the input prompt, unable to mix both within a single output. Bark \footnote{\url{https://github.com/suno-ai/bark}} attempts to mix speech and singing generation and is closest to achieving the goal of SCS; however, it relies on \textit{explicit tags} to control switching, lacking semantic awareness and also suffering from unstable performance. Therefore, existing models cannot facilitate automatic speech-singing switching driven solely by text content. \wen{I revised this part, as I thought Bark is also a unified audio generation framework, so that it is clear we discussed the limitations of various unified audio generation frameworks in achieving the goal of SCS.}

\wen{Yufei, please follow the suggestions in Abstract to provide a clear list of our innovations in the training paradigm, the dataset, and the CoT plan-then-generate approach for enhancing prosodic and melodic controllability.}
To address this gap, we introduce \textit{UniVocal}, a unified framework for speech-singing code-switching synthesis. Our approach adapts CosyVoice 2 \citep{du2024cosyvoice2}, a strong TTS model, through a data-efficient two-stage curriculum learning strategy. The stage-1 aligns speech and singing representations within a unified latent space, while  stage-2 develops automatic switching capability. A critical innovation in our framework is addressing the scarcity of SCS training data. We introduce a scalable pipeline to synthesize diverse code-switching data that is both semantically and acoustically natural. Specifically, we leverage LLMs to generate semantically coherent scripts across multiple scenarios that incorporate diverse switching triggers, ranging from explicit transitional phrases to implicit semantic cues. These scripts are then synthesized using our stage-1 model to ensure consistent acoustic quality, enabling the model to effectively learn semantic triggers for mode transitions.

Beyond the core SCS capability, we address the acoustic limitations of semantic tokenizers, which often discard fine-grained prosodic information. We integrate a Chain-of-Thought (CoT) approach that explicitly models pitch information via interleaved  refined cent tokens and semantic tokens. By defining the refined cent token as a high-resolution pitch representation, this design enables the model to plan prosody before content generation, enhancing prosody in speech and melody in singing. This "plan-then-generate" mechanism not only enhances melody in singing but also naturally improves the model's textual empathy capabilities, enabling richer emotional expression in speech.

\wen{Simiarly, please follow the suggestions in Abstract to summarize the test sets for evaluations, the main results and main findings.}
Experimental results demonstrate that \textit{UniVocal} achieves state-of-the-art mode-switching F1 scores (0.871 objective, 0.810 subjective) on \textbf{SCSBench-Mixed}, our code-switching speech-singing test set, outperforming cascaded baselines. Furthermore, the model also enhances empathetic expression while maintaining competitive performance on regular speech and singing generation benchmarks.

Our contributions can be summarized as follows:
\begin{itemize}[leftmargin=*,noitemsep]
\item We define the task of \textbf{Speech-Singing Code-Switching (SCS) Synthesis} and propose \textbf{UniVocal}, a unified framework that, guided by a global instruction defining the task scope, automatically infers vocal modes from textual semantics without explicit tags.
\item We introduce a \textbf{scalable data synthesis pipeline} and a \textbf{two-stage curriculum learning strategy} to address data scarcity. By constructing semantically and acoustically natural synthetic data with diverse scenarios and cues, we enable the model to master SCS capabilities efficiently.
\item We introduce a \textbf{refined cent token} within a CoT approach to mitigate the acoustic loss of semantic tokenizers. This explicitly models pitch information, enhancing prosodic planning and unlocking latent textual empathy capabilities.
\item We construct \textbf{SCSBench}, a comprehensive code-switching test set covering multiple cue types and scenarios. Experiments demonstrate state-of-the-art performance on \textbf{SCSBench} while maintaining competitive results on regular generation benchmarks.
\end{itemize}

\section{Related Work}
\noindent\paragraph{Task-specific Audio Generation} 
Audio generation has traditionally been divided into specialized domains with distinct architectures and training objectives. TTS models, such as Seed-TTS \citep{anastassiou2024seed} and CosyVoice \citep{du2024cosyvoice,du2024cosyvoice2} excel in speaker similarity and stability, but their training objectives limit them to spoken prosody patterns, preventing melodic expression. Conversely, music generation \citep{copet2023musicgen,yuan2025yue} and SVS \citep{zhang2022visinger,wang2024prompt} models are designed to follow explicit musical scores or structured lyrics, prioritizing musical accuracy over natural linguistic expression \citep{pan2025svs-survery}. This fundamental design difference creates a structural barrier: TTS models lack the melodic modeling capacity required for singing, while music/SVS models lack the prosodic flexibility needed for natural speech. As a result, these specialized approaches cannot support automatic code-switching between speech and singing within a single generation model, required by the SCS task.

\noindent\paragraph{Unified Audio Generation}  
Recent efforts have sought to unify multiple audio generation tasks within single frameworks, yet they fail to achieve automatic code-switching. Bark\footnote{\url{https://github.com/suno-ai/bark}} attempts mixed-mode generation through explicit control tags, but this approach requires manual annotation of tags and also suffers from mode transition instability. More importantly, it lacks semantic awareness to determine switching points based on text content alone. Unified frameworks such as UniSyn \citep{lei2023unisyn} and UniAudio \citep{yang2023uniaudio} employ multi-task training to handle both speech and singing, but they generate only one mode per input based on instruction prompts, preventing intra-sequence switching. Vevo2 \citep{zhang2025vevo2} introduces intermediate prosody tokens to model both speech and singing, but it requires reference audio to determine the output vocal mode, making it unable to infer mode switches solely from textual semantics. Different from these works, the proposed UniVocal can automatically infer speech-singing vocal mode transitions based on text content, without explicit control signals.\wen{I revised this part. It is better to clearly describe the differences between our work and related works, which helps highlight our innovations.}

\noindent\paragraph{Prosody Modeling and Tokenization} \wen{I think you meant for discrete-token-based audio generation systems, it is essential for the audio tokens to encode acoustic details for expressive speech and singing generation.} Semantic tokenizers \citep{hsu2021hubert,du2024cosyvoice,du2024cosyvoice2} extract high-level linguistic content but discard acoustic details, resulting in flat prosody. Conversely, acoustic tokenizers \citep{defossez2022encodec} achieve high fidelity but mix timbre, content, and prosody, making controllability difficult. To bridge this gap, research has explored explicitly separating prosody. Approaches using more F0 information \citep{zhao2020f0,kharitonov2021pGSLM} have shown promise for speech but often lack the melodic precision required for singing. Vevo2 \citep{zhang2025vevo2} uses a chromagram-based tokenizer to model both speech and singing; however, the 12-semitone resolution of chromagrams is too coarse to capture the fine-grained prosody of natural speech. Also, above tokenizations require training a complete codec architecture, demanding additional computational and data resources. In contrast to these works, we introduce a refined cent token to supplement the fine-grained acoustic details that are often missing in semantic tokenizers. Furthermore, we employ a CoT strategy to explicitly model pitch information, effectively enhancing both speech prosody and singing melody.

\section{Methodology}
\begin{figure}[t]
    \centering
    \includegraphics[scale=0.31]{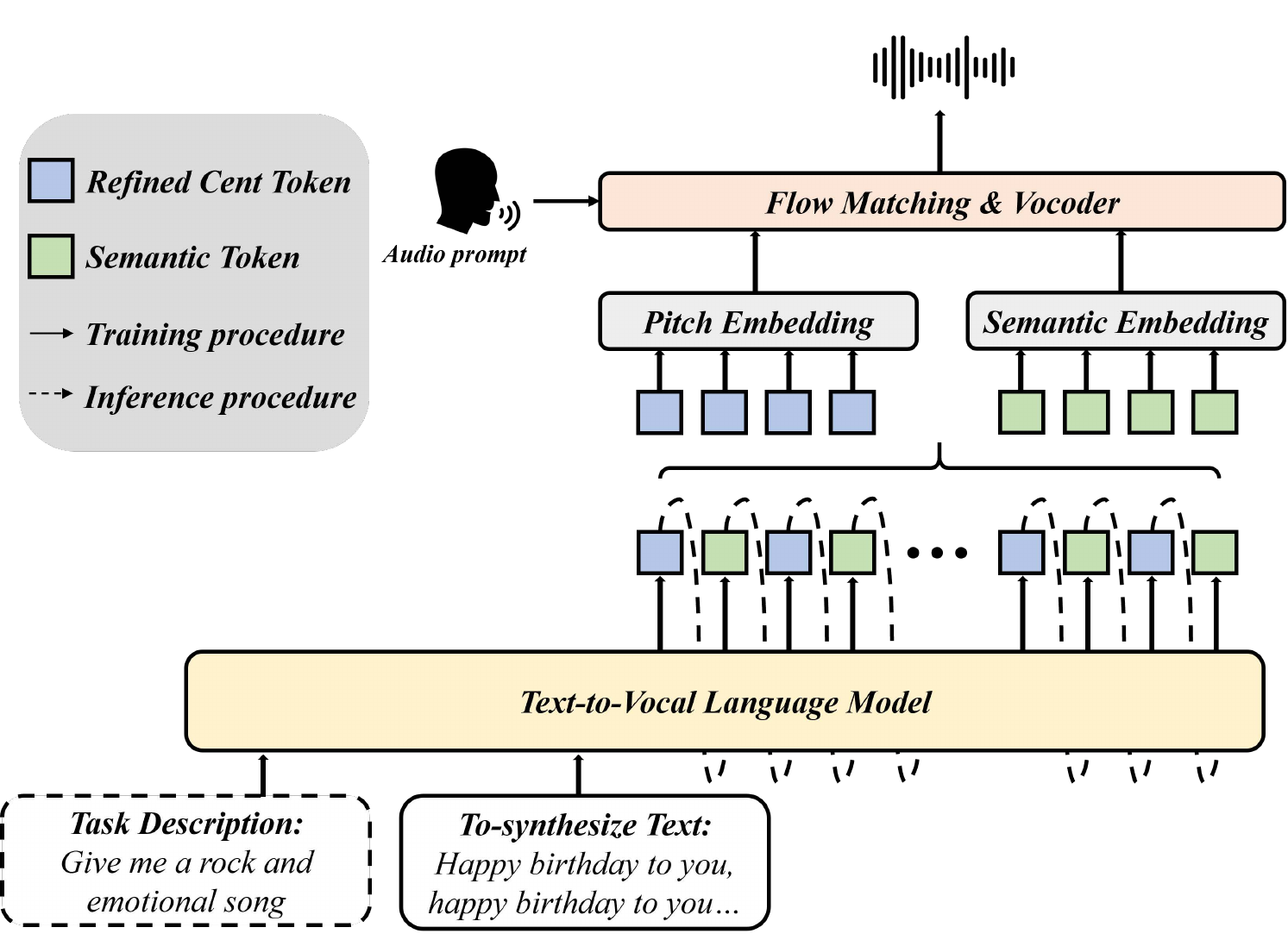}
    \caption{Overview of \textit{UniVocal}. The Text-to-Vocal language model receives the text to be generated, along with an optional natural language description of the task. At each timestep, it autoregressively generates a refined cent token and a semantic token in sequence. These two types of predicted tokens are then fed, along with the prompt audio, into a downstream module to synthesize the final voice output.}
    \label{fig:model}
\vspace{-15pt}
\end{figure}

\vspace{-5pt}
Figure \ref{fig:model} provides an overview of \textit{UniVocal}, which is a unified framework capable of executing diverse vocal generation tasks, including TTS, SVS, and SCS. We build upon semantic-token-based TTS architectures to prioritize generation stability. While our framework is model-agnostic, we employ CosyVoice 2 \citep{du2024cosyvoice2} as the backbone in this work due to its robust performance. To guide the model across these distinct tasks, we utilize global task-specific instructions. These instructions define the overall task scope, whereas the fine-grained speech-singing switching within the SCS task is inferred autonomously from text semantics. Furthermore, to address the inherent limitation of semantic tokens in capturing acoustic details, we introduce a refined cent token in a CoT approach. This design explicitly supplements the missing pitch information required for fine-grained speech prosody and singing melody.

\wen{Please remove ``training-free'' as they still need to be trained}\wen{(1) My understanding is that all the proposed methods, including the two-stage curriculum training, synthetic data generation, introduction of refined cent token, and the CoT method, are model-agnostic and can be applied to other strong TTS systems with semantic tokenizers. We just choose CosyVoice 2 as the backbone as it is a competitive open-source TTS system with xxxx advantages. So please make this clear and do not mislead the readers to consider our approach cannot be applied to other models. You could explain here we choose to adapt CosyVoice 2 in this work and emphasize the model-agnostic characteristics, then in the following text you could use CosyVoice 2 to describe the methodology. (2) Also, I think it helps to explain if we choose a dual-stream or hybrid-token TTS system that uses both semantic and acoustic tokens, do we still need the Refined Cent Token? Or a tradeoff between model simplicity, intelligibility, and naturalness, we choose to train a strong TTS system with semantic tokenizer, CosyVoice 2 in this work, and hence need to introduce the Refined Cent Token.}

\subsection{Refined Cent Token}
In contrast to raw $F_0$ in speech, music requires a perceptually meaningful representation \citep{huang2020spectral,tzanetakis2003pitch,gupta2017perceptual}. In western music theory, notes are organized into octaves, each containing 12 semitones with a frequency ratio of $2^{1/12}$. Conversely, while semitone scale works for symbolic music, it is too coarse for capturing speech prosody. To bridge this gap, we adopt the cent scale—dividing each semitone into 100 cents—to achieve high-resolution modeling for both speech prosody and singing melody.
We define the conversion from linear frequency $f_{Hz}$ to the logarithmic cent scale $f_{cent}$ as follows, using the standard A4 (440 Hz) as the reference pitch:
\vspace{-10pt}
\begin{equation}
f_{cent} = 1200 \times \log_2 \left( \frac{f_{Hz}}{440} \right),
\tag{1}\label{eq1}
\end{equation}
\vspace{-10pt}

To construct a unified token space that aligns with musical intervals, we project the absolute $f_{cent}$ into a single octave via a modulo operation. The discretized cent token $I(f_{cent})$ is formulated as:
\begin{equation}
\small
I(f_{cent}) = \begin{cases} 
    \lceil f_{cent} \pmod{1200} \rceil & \text{if } f_{Hz} \neq 0 \\ 
    -1 & \text{if } f_{Hz} = 0 
\end{cases}, 
\tag{2}\label{eq2}
\end{equation}

\noindent where $f_{Hz} = 0$ denotes unvoiced regions (assigned token -1), and the modulo 1200 operation wraps the pitch trajectory into a 1200-cent range (one octave). The ceiling operation $\lceil \cdot \rceil$ discretizes continuous cent values into integer tokens, introducing a maximum quantization error of 1 cent (approximately 0.08\% frequency deviation), which is perceptually negligible. This approach provides a scalable and fine-grained discrete representation that effectively supplements the semantic tokens with rich rhythmic and melodic information. The 1200-bin resolution meets the optimal granularity for the unique demands of \textit{UniVocal}, as validated by our ablation studies (Appendix \ref{app:CoT resolution}).

\subsection{Model Architecture}
The architecture of \textit{UniVocal}, as shown in Figure \ref{fig:model}, builds upon the CosyVoice 2 framework. The core backbone is a unified text-to-vocal language model (LM), implemented as a 24-layer causal Transformer with $\sim$0.5B parameters.

\noindent\textbf{Instruction-Driven Conditioning.} The original CosyVoice 2 features an instruction-following mechanism, where a natural language style description (suffixed with a special token \textit{``<|endofprompt|>''}) prefixes the input text to guide generation. We extend this paradigm by designing distinct task-specific instructions for \wen{also TTS?} the singing and SCS tasks (see Appendix \ref{app:instruction} for full templates), while retaining the default instruction-free format for regular TTS. These instructions act as global high-level control signals. Specifically for the SCS task, the instruction (such as \textit{``Generate a monologue.<|endofprompt|>''} sets the overall scenario, while the fine-grained switching between speech and singing is automatically driven by the content of the input text itself.

\noindent\textbf{Interleaved Chain-of-Thought Generation.} Unlike CosyVoice 2 that directly predicts semantic tokens, our LM autoregressively generates an interleaved sequence of refined cent tokens and semantic tokens (Figure~\ref{fig:model}). We expand the original LM vocabulary $V$ by appending the set of refined cent tokens $V_{cent}$ (comprising 1201 tokens: 1200 cent values plus the unvoiced token -1), initializing their embeddings randomly. We denote the input text conditioned by instructions as $\mathbf{X}$, and the target interleaved sequence of length $T$ as $\mathbf{Y} = \{(c_1, s_1), \dots, (c_T, s_T)\}$, where $c_t$ and $s_t$ represent the cent token and semantic token at step $t$, respectively. Inspired by CoT, we enforce a strictly sequential and interleaved generation order. Both token streams operate at a 25 Hz frame rate. For each frame $t$, the model first predicts the refined cent token $c_t$, which contains pitch information. This token is then appended to the history to explicitly guide the prediction of the subsequent semantic token $s_t$. The joint probability is factorized as:
{
\setlength{\abovedisplayskip}{10pt}
\setlength{\belowdisplayskip}{10pt}
\vspace{-5pt}
\begin{equation}
\small
P(\mathbf{Y}|\mathbf{X}) = \prod_{t=1}^{T} P(c_t | \mathbf{X}, \mathbf{Y}_{<t}) \cdot P(s_t | \mathbf{X}, \mathbf{Y}_{<t}, c_{t}).
\tag{3}
\end{equation}}
\vspace{-9pt}

During inference, we enforce this structure by applying logit masks: at cent-token-prediction steps, we mask out semantic tokens (setting their logits to $-\infty$) to ensure valid sampling from $V_{cent}$, and vice versa for semantic-token-prediction steps. This masking mechanism ensures strict following of the interleaved generation order. \textbf{This interleaved generation strategy makes the model first draft a structural pitch contour—effectively ``planning'' the prosodic and melodic framework—before generating the specific linguistic content and the remaining acoustic details}. As a modularized component, the refined cent token enables flexible configuration: it can be omitted to prioritize stability for \textit{alignment-heavy} tasks, or integrated to activate CoT prosodic planning for \textit{aesthetic-driven} scenarios.

\noindent\textbf{Waveform Reconstruction.} Following the LM stage, We modify the flow matching module of CosyVoice 2 by integrating a randomly initialized embedding layer to process the refined cent tokens as a supplementary condition. This adapted module generates Mel-spectrograms, which are subsequently reconstructed into waveforms via a pre-trained HiFi-GAN \citep{kong2020hifi} vocoder.

\subsection{Scalable Data Synthesis Pipeline}
\label{sec:data_pipeline}
To address the scarcity of SCS data, we introduce a scalable pipeline (detailed in Appendix \ref{app:code-switching dataset}) to synthesize diverse code-switching data that is both semantically and acoustically natural. The pipeline consists of three steps:

\noindent\textbf{Semantic Text Generation.} We leverage Gemini 2.5 Pro to generate naturally "boundary-blurring" scripts across diverse scenarios (such as monologues, podcasts). To ensure natural transitions, we design two types of triggers: (1) \textit{Implicit Cues}, which rely on the inherent semantic distinction between conversational prose and lyrical verses; and (2) \textit{Explicit Cues}, where transitional phrases (such as ``\textit{reminds me of a tune...}'') serve as semantic anchors. This ensures the switches are intrinsically driven by context rather than random splicing.

\noindent\textbf{Unified Acoustic Synthesis.} We utilize our stage-1 aligned model to synthesize the audio. Crucially, to maintain acoustic consistency, both speech and singing segments are conditioned on the same speaker embedding, eliminating timbre mismatches. Speech segments are further conditioned on emotion-specific reference audio to match the textual sentiment. Finally, the generated speech and singing segments are concatenated to form the complete code-switching samples.

\noindent\textbf{Quality Control.} Finally, we apply a filtering mechanism based on Word Error Rate (WER) to discard samples with severe articulation issues or misalignment, ensuring high-quality training data.

\subsection{Two-Stage Curriculum Learning}
To progressively equip the baseline model with expanded vocal generation capabilities, we implement a curriculum learning strategy, allowing the model to first bridge the distributional gap between modes and then master the complex task of automatic switching.

\noindent\textbf{Stage-1: Latent Representation Alignment.} We align speech and singing distributions within a unified latent space via continued pre-training on CosyVoice 2. We use a 4:1 singing-to-speech ratio (empirically determined to balance melodic learning and speech quality) and format data with task-specific instructions: singing includes style instructions\wen{Please detail the style instructions in Appendix}, speech uses standard format. This establishes independent generation capabilities for both modes.

\noindent\textbf{Stage-2: Autonomous Switching Learning.} In the SFT stage, we induce SCS capability using synthetic code-switching data\wen{How did we synthesize the code-switching data to ensure the code-switching is natural?}. To prevent catastrophic forgetting, we adopt a balanced 1:1:1 mixture of code-switching, speech, and singing data, ensuring competitive performance on regular tasks while mastering SCS.

\vspace{-5pt}
\section{Experimental Setup}
\subsection{Dataset}
Our training data spans three categories. 

\noindent\textbf{Code-Switching Data:} Following the pipeline in Section \ref{sec:data_pipeline}, we constructed 11,769 synthetic samples (262 hours) for the SCS task. 

\noindent\textbf{Speech Data:} We utilize 960 hours from LibriTTS \citep{zen2019libritts}, employing the full dataset for stage-1 and a 200-hour subset for stage-2. 
\noindent\textbf{Singing Data:} We curated 3,700 hours from Suno\footnote{\url{https://huggingface.co/datasets/nyuuzyou/suno}} (cleaning details in Appendix \ref{app:cleanning pipeline}) and included GTSinger \cite{zhang2024gtsinger}. We use the full Suno dataset for stage-1, while sampling a balanced 200-hour subset alongside 10 hours from GTSinger for stage-2.

\subsection{Training and Hyperparameters}
We build \textit{UniVocal} upon CosyVoice 2. The model is optimized via AdamW with a stage-dependent learning rate schedule. The entire two-stage training requires approximately 6 days on 4 NVIDIA A800 GPUs. Detailed hyperparameters and schedules are provided in Appendix \ref{app:training_details}.
\subsection{Evaluation Methodology}
\subsubsection{Evaluation Datasets} 
We evaluate \textit{UniVocal} across three distinct capability domains. More details of evaluation settings are provided in Appendix \ref{app:evaluation settings for each task}.

\noindent\textbf{Speech-Singing Code-Switching (SCS) Synthesis:} To assess mode-switching capabilities, we construct \textbf{SCSBench} as a held-out subset of the synthetic data generated in Section \ref{sec:data_pipeline}. It is stratified into: \textit{(1)} \textbf{SCSBench-Implicit} containing exclusively implicit semantic cues; \textit{(2)} \textbf{SCSBench-Explicit} containing explicit trigger phrases cues; and \textit{(3)} \textbf{SCSBench-Mixed} incorporating both cue types. This stratification allows us to analyze the model's sensitivity under different conditions.

\noindent\textbf{Single-Mode Tasks:} We evaluate regular and expressive TTS using the SeedTTS test set \citep{anastassiou2024seed} and a curated textual empathy test set, respectively. For singing, we utilize a held-out GTSinger \citep{zhang2024gtsinger} subset for short phrases and a constructed \textit{fullsong} set for long-form assessment.

\begin{figure}[t] 
    \centering
    \includegraphics[scale=0.31]{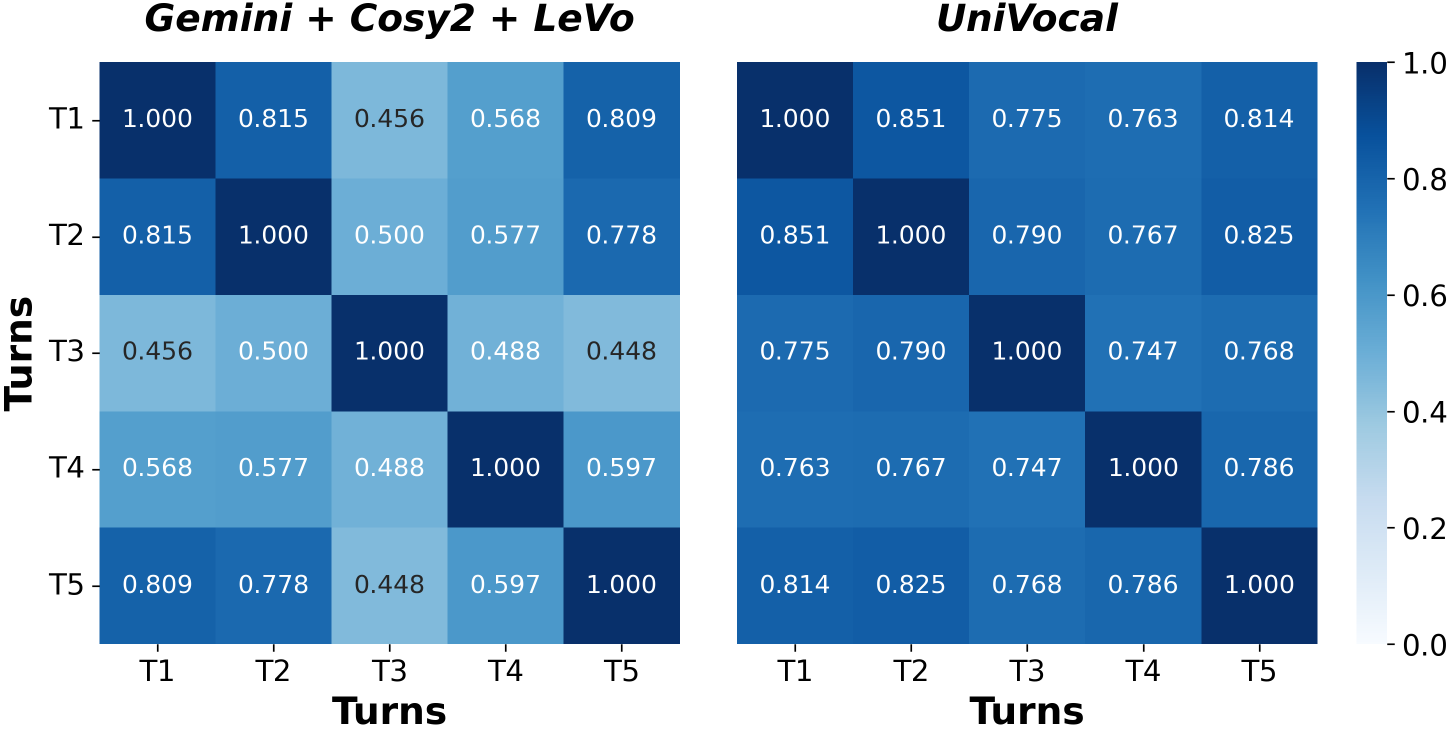}

    \vspace{-5pt} 
    \caption{ Intra-sample speaker consistency. Pairwise similarity heatmap between five temporal segments, averaged across all generated samples from each system. Darker colors indicate higher speaker stability.}
    \label{fig:sim_of_turns}
    \vspace{-5pt} 
\end{figure}

\vspace{-5pt}
\subsubsection{Evaluation Metrics}
We employ a comprehensive suite of objective and subjective metrics. Detailed definitions, implementation tools, and validation procedures are provided in Appendix \ref{app:f1 and MOS criteria} and \ref{app:f1_calibration}.

\noindent\textbf{Objective Metrics.} We follow standard evaluation protocols \citep{du2024cosyvoice2,lei2025levo}. For regular TTS, we report WER (semantic consistency) using Whisper-v3 \citep{radford2023whisper}, SIM (speaker similarity) using ERes2Net \citep{chen2023Eres2net}, and  UTMOS \citep{saeki2022utmos} (naturalness). For singing generation, we evaluate WER, SIM, AES \citep{tjandra2025meta} (aesthetic quality, averaged from CE, CU, PC, and PQ), and QUA (audio quality, averaged from OVRL and SRMR) using ClearVoice\footnote{\url{https://github.com/modelscope/ClearerVoice-Studio/tree/main/speechscore}}. For SCS transition accuracy, we report F1-scores based on segment-level singing mode identification by Gemini 2.5 Pro via In-Context Learning.


\noindent\textbf{Subjective Metrics.} Human annotators provide ratings using the 3-point scale: (1) E-MOS (empathy) and P-MOS (prosody) for empathetic TTS; (2) M-MOS (musicality) and N-MOS (naturalness) for singing; and (3) human-labeled F1-scores for SCS. Inter-annotator agreement was substantial ($\kappa=0.684$).

\subsubsection{Model Configuration}
\textit{UniVocal} offers two configurations: standard (omitting refined cent token) for alignment-heavy tasks (SCS, TTS), and expressive (with CoT) for aesthetic tasks (empathetic TTS, singing). All reported results use the optimal configuration per domain.

\vspace{-7pt}
\section{Results}
\vspace{-5pt}
\subsection{Main Results}
\label{sec:main result}
\noindent\textbf{Seamless Mode Switching.} We evaluate automatic mode-switching against two constructed cascaded baselines: \textit{Gemini + Bark} and \textit{Gemini + Cosy2 + LeVo}, which utilize Gemini 2.5 Pro for explicit text segmentation (see \ref{app:SCS evaluate details} for baseline details). As shown in Tables \ref{table:mix_results_f1} and \ref{table:mix_results_quality}, \textit{UniVocal} achieves state-of-the-art F1 scores across SCSBench, surpassing these strong cascaded systems on most subsets. On \textbf{SCSBench-Mixed}, it attains F1(O) of 0.871 and F1(S) of 0.810, proving its ability to infer mode switching when given sufficient semantic cues. Our model also achieves the lowest WER and highest UTMOS globally, reflecting superior content consistency and naturalness. While speaker similarity (SIM) slightly lags behind the \textit{Gemini + Cosy2 + LeVo} baseline—likely due to poor quality of the singing training data—our speaker similarity stability analysis shows different results. Adapted from inter-turn speaker consistency metrics in dialogue generation \citep{zhang2025covomix2}, we partition samples into five temporal segments to compute the average pairwise speaker similarity (Figure \ref{fig:sim_of_turns}). Results confirm \textit{UniVocal} maintains significantly higher identity consistency across switches than cascaded baseline, minimizing timbre drift.

\begin{table*}[t]
  \centering
  \small
  \caption{Mode-switching accuracy on SCSBench. F1(O) and F1(S) represent metrics evaluated by Gemini 2.5 Pro and human annotators, respectively, assessing the accuracy of speech-singing transition timing. The \textbf{bold} and \underline{underlined} numbers indicate the optimal and sub-optimal results, respectively.}
  \label{table:mix_results_f1}
  \vspace{-3pt}
  \begin{tabular}{l c c c c c c c} 
   \toprule
   \multirow{2}{*}{\textbf{Model}} & \multicolumn{2}{c}{\textbf{SCSBench-Implicit}} & \multicolumn{2}{c}{\textbf{SCSBench-Explicit}} & \multicolumn{2}{c}{\textbf{SCSBench-Mixed}} \\
   \cmidrule(lr){2-3} \cmidrule(lr){4-5} \cmidrule(lr){6-7}
    & F1(O) & F1(S) & F1(O) & F1(S) & F1(O) & F1(S) \\
   \midrule
    Gemini + Bark & 0.414 & 0.142 & 0.533 & 0.250 & 0.465 & 0.199 \\
    Gemini + Cosy2 + LeVo & \textbf{0.752} & \textbf{0.685} & \underline{0.572} & \underline{0.489} & \underline{0.607} & \underline{0.566} \\
    \textit{UniVocal} & \underline{0.626} & \underline{0.595} & \textbf{0.714} & \textbf{0.635} & \textbf{0.871} & \textbf{0.810} \\
   \bottomrule
  \end{tabular}
 \end{table*}
 
 \begin{table*}[t]
  \centering
  \begin{threeparttable}
  \small
  \caption{Speech quality metrics on SCSBench. The \textbf{bold} and \underline{underlined} numbers indicate the optimal and sub-optimal results, respectively.}
  \label{table:mix_results_quality}
  \vspace{-3pt}
  \begin{tabular}{l c c c c c c c c c c} 
   \toprule
   \multirow{2}{*}{\textbf{Model}} & \multicolumn{3}{c}{\textbf{SCSBench-Implicit}} & \multicolumn{3}{c}{\textbf{SCSBench-Explicit}} & \multicolumn{3}{c}{\textbf{SCSBench-Mixed}} \\
   \cmidrule(lr){2-4} \cmidrule(lr){5-7} \cmidrule(lr){8-10}
    & WER$\downarrow$ & SIM$\uparrow$ & UTMOS$\uparrow$ & WER$\downarrow$ & SIM$\uparrow$ & UTMOS$\uparrow$ & WER$\downarrow$ & SIM$\uparrow$ & UTMOS$\uparrow$ \\
   \midrule
    Gemini + Bark\tnote{1} & 21.83 & --- & 3.41 & 29.47 & --- & 3.31 & 29.60 & --- & 3.31 \\
    Gemini + Cosy2 + LeVo & \underline{17.97} & \textbf{0.758} & \underline{3.42} & \underline{8.18} & \textbf{0.763} & \underline{3.62} & \underline{12.43} & \textbf{0.773} & \underline{3.54} \\
    \textit{UniVocal} & \textbf{5.83} & \underline{0.650} & \textbf{4.36} & \textbf{8.80} & \underline{0.643} & \textbf{4.41} & \textbf{10.90} & \underline{0.652} & \textbf{4.36} \\
   \bottomrule
  \end{tabular}
  \begin{tablenotes}
            \footnotesize
            \item[1] Since Bark is limited to using fixed registered voices, we did not calculate SIM metrics for this baseline.
        \end{tablenotes}
    \end{threeparttable}
  \vspace{-3pt}
 \end{table*}

\noindent\textbf{Expressive Speech Preservation.} 
\textit{UniVocal} maintains competitive zero-shot TTS performance (Table \ref{table:speech_results}), ranking first in UTMOS on SeedTTS-EN with minimal SIM degradation. Crucially, on the empathy task, it significantly outperforms the CosyVoice 2 baseline (E-MOS/P-MOS +0.48), achieving emotional consistency comparable to commercial systems like ElevenLabs, thus confirming the unlocking of latent empathetic capabilities.

\begin{table}[t]
\centering
\begin{threeparttable}
\small
\vspace{-5pt}
\caption{Performance on zero-shot TTS and textual empathy tasks. The \textbf{bold} and \underline{underlined} numbers indicate the optimal and sub-optimal results.}
\label{table:speech_results}
\vspace{-3pt}
\begin{tabular}{l c c c} 
\toprule
\multicolumn{4}{c}{\textbf{(a) SeedTTS-EN}} \\
\cmidrule(lr){1-4}
\textbf{Model} & WER$\downarrow$ & SIM$\uparrow$ & UTMOS$\uparrow$ \\
\midrule
F5-TTS & \textbf{2.15} & \textbf{0.755} & 3.68 \\
CosyVoice 2 & 2.96 & \underline{0.744} & \underline{4.18} \\
Vevo 1.5 & 12.58 & 0.718 & 3.68 \\
\midrule
\textit{UniVocal} & \underline{2.69} & 0.703 & \textbf{4.21} \\

\midrule[\heavyrulewidth] 

\multicolumn{4}{c}{\textbf{(b) Textual Empathy Test set}} \\
\cmidrule(lr){1-4}
\textbf{Model} & E-MOS$\uparrow$ & P-MOS$\uparrow$ & WER$\downarrow$ \\
\midrule
CosyVoice 2 & 1.78 & 1.74 & 0.53 \\
Multilingual-v2\tnote{1} & \textbf{2.30} & \textbf{2.47} & \textbf{0.24} \\
\midrule
\textit{UniVocal} & \underline{2.26} & \underline{2.22} & \underline{0.32} \\
\bottomrule
\end{tabular}
\begin{tablenotes}
            \footnotesize
            \item[1] Refers to ElevenLabs’
multilingual-v2.
        \end{tablenotes}
    \end{threeparttable}
\vspace{-15pt}
\end{table}

\noindent\textbf{Balancing Melodiousness and Quality.} 
For singing (Table \ref{table:singing_results}), \textit{UniVocal} ranks first in WER and QUA on GTsinger. While objective metrics on the \textit{fullsong} set are consistently high, subjective evaluations provide the definitive validation: \textit{UniVocal} surpasses Vevo 1.5 in both naturalness (2.23 vs 2.17 N-MOS) and musicality (2.18 vs 2.08 M-MOS), successfully balancing vocal fidelity with melodic constraints.

\begin{table*}[t]
 \centering
 \begin{threeparttable}
 \small
\renewcommand{\arraystretch}{1.1}
 \caption{Results on the singing generation task. The \textbf{bold} and \underline{underlined} numbers indicate the optimal and sub-optimal results, respectively.}
 \label{table:singing_results}
 \vspace{-3pt}
 \begin{tabular}{l c c c c c c c c c c} 
  \toprule
  \multirow{2}{*}{\textbf{Model}} & \multicolumn{4}{c}{\textbf{GTsinger}}&\multicolumn{6}{c}{\textbf{Fullsong}}\\
  \cmidrule(lr){2-5} \cmidrule(lr){6-11}
  &AES$\uparrow$ & WER$\downarrow$ & SIM$\uparrow$ & QUA$\uparrow$&
  AES$\uparrow$ & WER$\downarrow$ & SIM$\uparrow$ & QUA$\uparrow$& N-MOS$\uparrow$ & M-MOS$\uparrow$\\
  \midrule
    Vevo 1.5 & \underline{5.38} &\underline{22.79} & \textbf{0.709} & 8.71&\underline{5.46} &\underline{49.55} & \underline{0.66} & 6.97 & 2.17 & 2.08\\
    YuE\tnote{1}  & 5.32 & 40.32 & 0.352 & \underline{9.33} & 5.14& 77.60 & 0.46 & 6.34 & 2.22 & \underline{2.24}\\
    LeVo\tnote{2}  & 5.25 & 23.44 & 0.603 & 9.30& 5.37 & 69.41 & 0.54 & \underline{7.51}  & \textbf{2.41} & \textbf{2.34}\\
   \midrule
    \textit{UniVocal} & \textbf{5.44} & \textbf{18.07} & \underline{0.703} &\textbf{10.70}& \textbf{5.58} & \textbf{35.88} & \textbf{0.72} & \textbf{7.75} & \underline{2.23} & 2.18  \\
  \bottomrule
 \end{tabular}
 \begin{tablenotes}
            \footnotesize
            \item[1, 2] For LeVo and YuE, metrics are computed on vocal tracks.
        \end{tablenotes}
        \vspace{-5pt}
    \end{threeparttable}
\end{table*}

\vspace{-5pt}
\subsection{Ablation Studies}
To comprehensively evaluate our framework, we examine the contributions of both the modular refined cent token and the curriculum learning strategy. We compare the expressive configuration (\textit{UniVocal}) against the standard configuration (denoted as \textit{w/o CoT}) and \textit{UniVocal} trained without the two-stage curriculum (denoted as \textit{w/o CL}). Table \ref{table:ablation} presents the results of these variants.

\begin{table*}[t]
 \centering
 \small
 \renewcommand{\arraystretch}{1.2}
 \caption{Ablation studies on the proposed framework. We report results for the expressive configuration (\textit{UniVocal}), and variants removing the refined cent token (\textit{w/o CoT}) or the curriculum learning strategy (\textit{w/o CL}). The \textbf{bold} and \underline{underlined} numbers indicate the optimal and sub-optimal results, respectively.}
 \label{table:ablation}
 \vspace{-3pt}
 \begin{tabular}{l c c c c c c c c} 
  \toprule
  \multirow{2}{*}{\textbf{Model}} & \multicolumn{3}{c}{\textbf{Textual Empathy Test Set}} & \multicolumn{3}{c}{\textbf{\textbf{Fullsong}}} & \multicolumn{2}{c}{\textbf{SCSBench-Mixed}} \\
  \cmidrule(lr){2-4} \cmidrule(lr){5-7} \cmidrule(lr){8-9}
   & E-MOS$\uparrow$ & P-MOS$\uparrow$ & WER$\downarrow$ & N-MOS$\uparrow$ & M-MOS$\uparrow$ & WER$\downarrow$ & F1$\uparrow$ & WER$\downarrow$ \\  
  \midrule
  \textit{UniVocal} & \textbf{2.26} & \underline{2.22} & \textbf{0.32} & \underline{2.23} & \textbf{2.18} & \textbf{35.30} & \underline{0.716} & \textbf{5.99} \\
  ~~\textit{w/o CoT} & 2.03 & 1.84 & \underline{0.51} & 2.20 & 1.86 & \underline{35.88} & \textbf{0.810} & \underline{10.90} \\
  ~~\textit{w/o CL} & \underline{2.24} & \textbf{2.23} & 0.52 & \textbf{2.29} & \underline{2.17} & 37.21 & 0.496 & 14.46 \\
  \bottomrule
 \end{tabular}
 \vspace{-5pt}
\end{table*}

\noindent\textbf{Impact of CoT.} Comparing the expressive configuration with the standard configuration (\textit{w/o CoT}) reveals a clear trade-off. While removing CoT yields marginal gains in switching stability (SCS F1 increases to 0.810), it comes at the cost of intelligibility (WER increases) and aesthetic performance (with E-MOS and P-MOS decreasing by 0.23 and 0.38, and N-MOS and M-MOS dropping by 0.03 and 0.32, respectively.). Without the pitch planning from CoT, the model suffers a substantial drop in subjective MOS scores, confirming that CoT is essential for highly expressive generation.
Correlations between predicted and ground-truth cent tokens confirm this mechanism actively drafts a structural pitch framework rather than merely supplementing acoustic details (Appendix \ref{app:pitch plan}).

\noindent\textbf{Necessity of Curriculum Learning Strategy.} To validate the effectiveness of our progressive adaptation strategy, we trained a variant by jointly optimizing on the full mixture of speech, singing, and synthetic code-switching data (denoted as \textit{w/o CL}), effectively bypassing the stage-1 alignment. As shown in Table \ref{table:ablation}, this one-stage approach leads to suboptimal convergence. Although the model retains reasonable TTS capabilities, it struggles to capture the subtle semantic triggers for mode-switching (F1 drops to 0.496). This indicates that the latent space alignment established in stage-1 is a prerequisite for effectively mastering the complex SCS task.

\noindent\textbf{The Source of Empathy Capabilities.} 
CoT is not the sole driver of empathy. The \textit{w/o CoT} variant already outperforms the baseline (E-MOS 2.03 vs 1.78). To isolate the source of this gain, we trained a supplementary variant using only stage-1 data without CoT, which yielded similar E-MOS scores to \textit{w/o CoT}. This confirms that stage-2 data has negligible impact. We therefore attribute empathy to a synergy: emotionally diverse singing data in stage-1 unlocks latent expressive representations, which are further amplified by the refined cent token.

\subsection{Qualitative Analysis on Switching Cues} 
\label{sec:case_study}
To investigate the model's sensitivity to semantic triggers, we conduct a case study across different types of cues, as defined in Appendix \ref{apps:cues definition}: implicit cues refer to the inherent semantic differences between speech and singing, while explicit cues are transitional trigger phrases inserted before singing parts. Table \ref{tab:case_study} presents the inference outcomes.

\begin{table}[t] 
    \centering
    \footnotesize 
    \setlength{\tabcolsep}{3pt} 
    \renewcommand{\tabularxcolumn}[1]{m{#1}}
    \caption{Case study on mode-switching under different cue types. \textbf{Bold} and \textit{italics} denote expected and successfully generated singing parts, respectively. Implicit Only features semantic distinctions between modes, while Explicit Only uses \textcolor{red}{red} trigger phrases with minimized semantic disparity.}
    \label{tab:case_study}
    \vspace{-6pt}
    \begin{tabularx}{\linewidth}{l X c}
    \toprule
    \textbf{Cue Type} & \textbf{Example} & \textbf{Outcome} \\
    \midrule
    \makecell[l]{Explicit\\+ Implicit} 
    & He begin to sing softly, \textcolor{red}{always the same tune}. \textbf{\textit{Oh, the river flows, and the wild wind blows... }} He'd trail off... 
    & Accurate \\
    \midrule
    Explicit Only 
    & There's a lyric that really fits this mood. \textcolor{red}{It goes...} \textbf{\textit{the streetlights look the same from every window.}} And they really do.  
    & Accurate \\
    \midrule
    Implicit Only 
    & I wasn't just in my kitchen making coffee. \textbf{\textit{Mmm-hmm, mm-mm...}} It's amazing how a few notes can do that. \textbf{We were young and free, just you and me.} And for a second... 
    & \makecell{Partial\\Failure} \\ 
    \bottomrule
    \end{tabularx}
\vspace{-10pt}
\end{table}

\noindent\textbf{Impact of Explicit Anchors.} As observed in the first two rows, the presence of explicit cues—whether combined with implicit semantic difference or used alone—significantly improves the switching success. Phrases like \textit{``always the same tune''} and \textit{``It goes...''} serve as strong anchors, effectively priming the model to switch modes at the correct boundary.

\noindent\textbf{Challenges in Implicit-Only Scenarios.} In the absence of explicit triggers, the model relies solely on semantic inference, which proves challenging. As shown in the ``Implicit Only'' case, the model fails to switch for the lyrical line \textit{``We were young and free...''}, likely misinterpreting it as narrative prose due to the lack of structural semantic difference from speech parts.

\noindent\textbf{The ``Humming'' Exception.} Interestingly, the same implicit-only sample demonstrates a successful switch on the humming segment (\textit{``Mmm-hmm, mm-mm''}). Although classified as an implicit cue, humming possesses a unique non-lexical textual form that contrasts sharply with speech. This distinct feature acts as a ``strong'' implicit cue, allowing the model to robustly generate singing prosody even without explicit triggers.
\vspace{-5pt}
\section{Conclusion}
\vspace{-5pt}
We introduce \textbf{UniVocal}, a unified framework that pioneers \textbf{Speech-Singing Code-Switching (SCS) Synthesis} by autonomously inferring vocal modes from text. We address data scarcity via a scalable synthesis pipeline and enhance acoustic modeling through a refined cent token with CoT planning. \textit{UniVocal} achieves state-of-the-art performance on \textbf{SCSBench} and maintaining competitive performance on regular generation benchmarks.
\section*{Limitations}
\noindent\textbf{Data Quality Constraints.} Our singing training data relies on synthetic songs generated by Suno, which are processed via source separation and ASR tools. Due to the inherent limitations of the source audio and these processing pipelines, a significant portion of the data suffers from artifacts (typified by electronic tones) and lyric misalignment. These imperfections inevitably impose an upper bound on the acoustic fidelity and semantic consistency of the singing segments generated by \textit{UniVocal}.

\noindent\textbf{Gap in Realistic Scenarios.}  A distributional gap remains between our synthetic training data and complex, real-world SCS scenarios. Consequently, \textit{UniVocal} currently relies on minor explicit semantic triggers to achieve robust generalization in natural settings (detailed in Appendix \ref{app:real SCS}), highlighting an area for future improvement in handling purely implicit transitions.

\noindent\textbf{Evaluation Precision.} 
While ICL strategy aligns Gemini 2.5 Pro with human preference at the system level (achieving perfect rank consistency), we acknowledge limitations in statistical resolution at the sample level. 
As detailed in Appendix \ref{app:f1_calibration}, the magnitude of sample-level correlation coefficients is inherently dampened by the discrete nature of F1 scores on short samples. Since the generated audio segments are often brief, the F1 metric frequently collapses into binary outcomes (0.0 or 1.0), lacking the continuous variance required for high linear correlation metrics. Consequently, while the automated metric serves as a reliable and scalable proxy for system-level benchmarking, its sensitivity to subtle, non-binary quality variations in individual short samples remains limited compared to fine-grained human perception.

\section*{Ethical considerations}
We acknowledge that the generation capabilities of \textit{UniVocal} could potentially be misused for deepfakes. We emphasize that this research is intended solely for academic purposes. Regarding data compliance, we utilized the open-sourced LibriTTS and Suno datasets, avoiding copyright disputes and privacy concerns associated with real-world vocal recordings. To mitigate risks, we release our models under a restrictive license that strictly prohibits commercial misuse and illegal impersonation.



\bibliography{custom}

\appendix
\section{Details of Dataset}

\label{app:dataset}
\subsection{Code-Switching Speech-Singing Dataset}
\label{app:code-switching dataset}

\subsubsection{Overview}
To enable the model to learn autonomous mode switching, we constructed a synthetic code-switching speech-singing dataset comprising 11,769 samples (262 hours). The dataset covers three scenarios—monologue, personal podcast, and audiobook—spanning diverse emotions. Structurally, each sample consists of a spoken narrative naturally interwoven with sung or hummed phrases. We utilize 9 distinct speaker timbres selected from diverse voice characteristics to maximize acoustic diversity. Key statistics of the dataset are summarized in Table \ref{table:dataset_key_information}. A representative subset of 1,200 samples (approximately 10\%) is reserved as \textbf{SCSBench} for evaluation, maintaining the roughly same scenario distribution as the training set.

\subsubsection{Construction Pipeline}
\label{apps:cues definition}
The dataset construction follows a three-step pipeline designed to simulate ``boundary-blurring'' switches:

\noindent\textbf{Semantic Text Generation:} To facilitate the construction of naturally ``boundary-blurring'' switching scripts, we employ Gemini 2.5 Pro~\citep{comanici2025gemini} to generate content where vocal mode switches are intrinsically driven by semantic context. Designed for first-person scenarios such as monologues, podcasts, and audiobooks, the scripts are primarily spoken but feature naturally inserted singing or humming segments. To enable \textit{UniVocal} to autonomously infer mode transitions solely from text, we enforce strict semantic distinctiveness between modes via a constrained prompting strategy. This establishes \textbf{implicit cues}: speech is maintained in a natural, prose-like conversational tone to drive narrative and logic, whereas singing is characterized by lyrical, repetitive, and emotionally heightened patterns. Humming segments (such as \textit{``hmm hmm hmm''}) utilize inherently unique textual representations. To further enhance robustness, we insert \textbf{explicit cues} (transitional trigger phrases such as \textit{``And that reminds me of a lyric...''} or \textit{``Let me try to recall that jingle for you...''}) immediately preceding singing parts in approximately 50\% of the samples. These phrases, expanded by Gemini 2.5 Pro from seed examples, serve as strong semantic anchors to stabilize the learning of speech-singing code-switching. All descriptive words containing special tags in scripts are excluded.  All generated text content is then compiled into JSON files.


\noindent\textbf{Audio Synthesis:} We utilize our stage-1 aligned model to generate segments for both modes. We apply differential conditioning: speech segments are synthesized conditioned on emotion-specific reference audio (sourced from Expresso \citep{nguyen2023expresso} and EmoVoice-DB \citep{yang2025emovoice} datasets) to align with the textual sentiment. This reference audio set comprises prompts from 9 speakers covering nine emotions: \textit{confused, happy, sad, angry, surprised, fearful, disgusted, default, laughing}. In contrast, singing segments are conditioned solely on the target speaker embedding (injected during the Flow Matching inference stage). These segments are subsequently concatenated---with a 0.25s silence interval inserted between segments---to form the complete code-switching audio samples. This approach preserves the specific emotional tone in speech while maintaining a consistent speaker identity across the concatenated sample. The emotion distribution of the resulting samples is skewed towards ``\textit{happy}'' (50\%), ``\textit{sadness}'' (16\%), and ``\textit{default}'' (11.6\%), with other emotional states collectively accounting for the remaining $\sim$12.4\%.

\noindent\textbf{Quality Filtering:} We filter samples based on Word Error Rate (WER) computed using Whisper-v3 \citep{radford2023whisper}. The 20\% threshold is empirically determined to balance data quality and quantity: samples with WER $\ge$ 20\% are discarded as they indicate severe misalignment, while those with moderate WER (10-20\%) are retained with ASR-transcribed text to ensure precise alignment. This filtering process removes approximately 15\% of initially generated samples. Finally, valid samples are formatted for training by prepending the scenario-specific instruction (such as \textit{``Generate a monologue''}) and the \textit{``<|endofprompt|>''} separator to the to\text{-}synthesize text.

\begin{table}[t]
 \centering
 \small
 \caption{Key statistics of the code-switching speech-singing dataset}
 \label{table:dataset_key_information}
 \begin{tabular}{l r r r} 
  \toprule
  \textbf{Scenario} & \textbf{Count} & \textbf{Total Dur. (h)} & \textbf{Avg. Dur. (s)} \\
  \midrule
   Monologue & 6,247 & 84.3 & 48.6 \\
   Podcast & 2,432 & 87.2 & 129.1 \\
   Audiobook & 3,090 & 90.4 & 105.3 \\
  \midrule
  \textbf{Sum} & \textbf{11,769} & \textbf{261.9} & \textbf{80.1} \\
  \bottomrule
 \end{tabular}
\end{table}

\subsection{Singing Data Cleaning Pipeline}
\label{app:cleanning pipeline}
To enable singing generation, we created an approximately 3,700-hour English singing dataset from the 23,000-hour open-source Suno\footnote{\url{https://huggingface.co/datasets/nyuuzyou/suno}} music dataset. Our objective was to extract only the English vocal tracks to mitigate training complexity. The process began with source separation using MelBand Roformer (viperx edition)\footnote{\url{https://github.com/ZFTurbo/Music-Source-Separation-Training/blob/main/docs/pretrained_models.md}}. A significant portion (60$\%$) of the separated audio was then filtered out using DNSMOS (OVRL) and SRMR metrics by ClearVoice-Studio\footnote{\url{https://github.com/modelscope/ClearerVoice-Studio/tree/main/speechscore}} to remove tracks with strong background noise and heavy reverberation. We applied MelBand Roformer (anvuew edition)\footnote{\url{https://github.com/ZFTurbo/Music-Source-Separation-Training/blob/main/docs/pretrained_models.md}} for dereverberation to reduce the electric tone. Next, we segmented the audio using an energy-based voice activity detection (VAD) method, merging adjacent segments into clips up to 4 minutes long. Lyrics were transcribed using FastWhisper \footnote{\url{https://huggingface.co/Systran/faster-whisper-large-v3}}. As FastWhisper's WER is high for singing, we followed RapBank's \citep{ning2025rapbank} methodology, calculating the phoneme-per-second (PPS) rate of the lyrics and discarding segments with rates that were too high or low to filter out hallucinations. Ultimately, we obtained approximately 3,700 hours of data. We then used Gemini 2.5 Pro to generate 100 natural language instruction templates (see Table \ref{table:singing_templates} for examples), which we populated with style tags from the original metadata (such as \textit{``A song featuring pop...<|endofprompt|>< to\text{-}synthesize text>''}). It is important to note that due to limitations in the metadata and separation tools, the quality of the final dataset remains suboptimal. This is primarily manifested as persistent, noticeable electric tone and a weak correlation between the style tags and the actual singing performance. We use the full singing data in the stage-1, whereas for the stage-2, we sample approximately 200 hours for training.

\subsection{Instruction Templates}
\label{app:instruction}

We design distinct instruction formats to condition the model for different generation tasks. The specific prompt structures and templates are detailed below.

\noindent\textbf{Speech-Singing Code-Switching (SCS) Synthesis.} As shown in Table \ref{tab:scs_prompts}, the instruction explicitly defines the narrative scenario for the SCS task. We employ three fixed instruction tokens corresponding to the dataset scenarios. During training and inference, one of these instructions prefixes the input text, separated by the special token ``\textit{<|endofprompt|>}''.

\begin{table}[h]
\centering
\small
\renewcommand{\arraystretch}{1.3}
\caption{Instruction templates for the SCS task. \textit{ Note: These prompts activate the SCS task globally. No segment-level tags (e.g.,<sing>, <speech>) are used within the input text.}}
\label{tab:scs_prompts}
\begin{tabular}{l l}
\toprule
\textbf{Scenario} & \textbf{Instruction Prompt} \\
\midrule
Monologue & \textit{Generate a monologue. <|endofprompt|>} \\
Podcast & \textit{Generate a podcast. <|endofprompt|>} \\
Audiobook & \textit{Generate an audiobook. <|endofprompt|>} \\
\bottomrule
\end{tabular}
\end{table}

\noindent\textbf{Speech Generation}
For regular speech synthesis tasks, no additional instruction prefix is required. The model takes the text content as input, following the default behavior of the baseline model.

\noindent\textbf{Singing Generation.} For singing tasks, the instruction is constructed to encapsulate the musical style. To ensure robustness and linguistic diversity, we utilized Gemini 2.5 Pro to generate 100 distinct natural language description templates containing placeholders for style tags. 

Formally, given the comprehensive metadata for a song, we employ a dynamic sampling strategy to enhance robustness. We randomly sample a subset of style tags $S$ (where $|S| = k$ and $1 \le k \le 5$) to fit the maximum capacity of our templates. Subsequently, we select a template $T$ specifically designed with $k$ placeholders to form the final instruction:
\begin{equation*}
\text{Input\_text} = T(S) \oplus \textit{<|endofprompt|>} \oplus \text{Lyrics}
\end{equation*}

Table \ref{table:singing_templates} lists a subset of these templates. The \texttt{\{style\}} placeholder is replaced by the selected style tags from the singing metadata during training.

\begin{table*}[t]
\centering
\small
\renewcommand{\arraystretch}{1.2}
\caption{Selected examples of the 100 natural language templates used for singing instructions. The \texttt{\{style\}} slot is dynamically filled with metadata tags.}
\label{table:singing_templates}
\begin{tabular}{p{0.95\linewidth}}
\toprule
\textbf{Singing Generation Instruction Templates (Selected)} \\
\midrule
\textit{Generate a song in the \texttt{\{style\}} style.<|endofprompt|>} \\
\textit{Sing a song with a \texttt{\{style\}} atmosphere. <|endofprompt|>} \\
\textit{I want a song that encapsulates \texttt{\{style\}}. <|endofprompt|>}\\
\textit{I need a \texttt{\{style\}} song with \texttt{\{style\}}. <|endofprompt|>} \\
\textit{A song that is \texttt{\{style\}}, featuring \texttt{\{style\}}. <|endofprompt|>} \\
\textit{Generate a \texttt{\{style\}} song, featuring \texttt{\{style\}} and \texttt{\{style\}}. <|endofprompt|>} \\
\textit{Create a track that's \texttt{\{style\}}, \texttt{\{style\}}, \texttt{\{style\}}, and \texttt{\{style\}} in style. <|endofprompt|>} \\
... \textbf{(100 templates in total)} \\
\bottomrule
\end{tabular}
\end{table*}

\section{Training Details}
\label{app:training_details}

\subsection{Hyperparameters}
The Language Model (LM) is optimized using AdamW with $\beta_1=0.9, \beta_2=0.95$, weight decay $0.1$, and gradient clipping at $1.0$. The learning rate schedules differ by stage:
\begin{itemize}
    \item \textbf{Stage-1:} Linear decay from $2 \times 10^{-4}$ to zero over 70,000 steps (with 5,000 warmup steps).
    \item \textbf{Stage-2:} Constant learning rate of $1 \times 10^{-4}$ for 30,000 steps.
\end{itemize}
Both stages utilize dynamic batching with a maximum capacity of 4.5 minutes per batch.

\subsection{Compute Resources}
All experiments were conducted on 4 NVIDIA A800 GPUs. Stage-1 training took approximately 5 days, while Stage-2 took 1 day. The Flow Matching module was fine-tuned on Stage-2 data for 2 days using a batch size of 2 minutes per GPU.

\section{Details of Evaluation}
\subsection{Evaluation Settings for Each Task}
\label{app:evaluation settings for each task}

This section details the evaluation dataset statistics, model inference configurations, and baseline implementations that supplement the methodology described in the main text.

\subsubsection{Speech-Singing Code-Switching (SCS)}
\label{app:SCS evaluate details}
\noindent\textbf{SCSBench Statistics.} The constructed \textbf{SCSBench} comprises approximately 1,210 samples. The dataset is strictly balanced across two dimensions:
\begin{enumerate}
    \item \textbf{Cue Types:} The three subsets (\textbf{SCSBench-Implicit}, \textbf{SCSBench-Explicit}, and \textbf{SCSBench-Mixed}) each constitute approximately one-third of the total data.
    \item \textbf{Scenarios:} Within each subset, samples are equally distributed across the three narrative scenarios: monologue, podcast, and audiobook.
\end{enumerate}

\noindent\textbf{Inference Configurations.} For \textit{UniVocal}, we generate audio using the task-specific instruction (as defined in Appendix \ref{app:instruction}) and a randomly assigned speaker for each sample. For the cascade baselines, we implement a multi-stage pipeline to enable code-switching:
\begin{itemize}
    \item \textit{Gemini + Bark:} We first employ Gemini 2.5 Pro to segment the input text into speech and singing parts based on semantic content. Since Bark has a maximum duration limit ($<$20s), we further segment long text parts using the NLTK\footnote{\url{https://pypi.org/project/nltk/}} library. After synthesizing each segment independently, we concatenate them with a 0.1s silence interval inserted between mode switches to form the final audio.
    \item \textit{Gemini + CosyVoice 2 + LeVo:} Similar to the above, we use CosyVoice 2 for speech segments and LeVo for singing segments. Both models are conditioned on the same target speaker to maintain identity consistency. Segments are concatenated with a 0.1s silence interval.
\end{itemize}

\noindent\textbf{Metric Calculation.} To evaluate mode-switching performance, we treat the presence of singing as the positive class and adopt a two-step evaluation procedure. First, at the sentence level, we determine the ground truth and prediction labels. Gemini 2.5 Pro and human annotators are instructed to classify each generated audio segment as speech or singing based on its acoustic characteristics. Second, based on these segment-wise classifications, we compute the metrics: \begin{itemize} \item Macro-F1: Calculated by computing the F1-score for each individual audio sample (sample-level) and then averaging these scores across the entire test set. \item Micro-F1: Calculated by aggregating all segment-level predictions (TP, FP, FN) globally across the dataset before computing the metric. \end{itemize}

\subsubsection{Textual Empathetic Speech}
The textual empathetic speech test set consists of 50 colloquial sentences designed to elicit emotional expressiveness. The dataset covers 10 distinct emotional scenarios, with 5 samples per category: \textit{anger, disgust, contempt, fear, hope, pride, joy, nostalgia, surprised,} and \textit{sadness}.

During inference, to strictly evaluate the model's ability to infer emotion from text content, all models are conditioned solely on speaker timbre without any emotion-specific reference audio. For the commercial baseline \textit{ElevenLabs multilingual-v2 model}, we utilize the ``Brian'' voice from official website\footnote{\url{https://elevenlabs.io/app/speech-synthesis/text-to-speech}}.

\subsubsection{Singing Generation}
We utilize two test sets to evaluate short-phrase and long-form singing generation capabilities.

\begin{itemize}
    \item \textbf{GTSinger (Short-phrase):} This set contains 200 lyrics samples. Each sample is paired with a reference audio prompt of 5--10 seconds. During inference, all models are conditioned on the style instruction for ``Pop'' music and perform in-context learning using the provided audio prompt.
    \item \textbf{Fullsong (Long-form):} This set comprises 27 tracks covering 9 musical styles: \textit{blues, country, electro, emotional, folk, jazz, pop, rap,} and \textit{rock}. Each sample is assigned a style-specific audio prompt (5--20 seconds) sampled from the training data. Models generate singing using the corresponding style instruction and the assigned prompt.
\end{itemize}

\paragraph{Metric Notes.}
Due to the limited style diversity in GTSinger, the MuQ-T metric is reported only for the \textit{fullsong} test set. For the FAD calculation, the reference distribution is constructed using the audio prompts from the respective test sets.

\subsection{Scoring Criteria and Protocols}
\label{app:f1 and MOS criteria}

This section details the human annotation setup, the specific logic for determining mode-switching accuracy (F1 score), and the granular scoring criteria used for both objective (Gemini-based) and subjective (human-based) evaluations.

\subsubsection{Human Annotator Configuration}
To ensure high-quality evaluation, we recruited paid participants to perform subjective assessments. The recruitment criteria were tailored to the specific tasks:
\begin{itemize}
    \item \textbf{Singing-Specific Criteria:} For singing generation task, we recruited annotators with professional musical backgrounds or formal musical training to ensure reliable judgment.
    \item \textbf{Other Task's Criteria:} All participants were required to demonstrate high proficiency in English listening comprehension for reliable judgment.
\end{itemize}
Each audio sample was evaluated by at least three independent annotators. Each audio sample was evaluated by at least three independent annotators. The reported scores represent the mean of these independent ratings. To validate the consistency of human judgments, we computed the inter-annotator agreement using Fleiss' kappa, yielding a score of $\kappa=0.684$, which indicates substantial agreement among annotators.

\begin{table*}[h]
    \centering
    \small
    \caption{Configuration of the Contrastive Few-Shot Examples used in ICL. The ``Hard Negative'' forces the model to ignore lyrical text when the acoustics indicate speech.}
    \label{tab:icl_examples}
    \begin{tabular}{l p{5cm} p{3.5cm} l}
    \toprule
    \textbf{Case Type} & \textbf{Audio Content Description} & \textbf{Textual Content (Lyrics)} & \textbf{Ground Truth Label} \\
    \midrule
    \textbf{Hard Negative} & \textbf{Spoken} recitation with natural speaking cadence & \textit{``We were both young when I first saw you...''} & \textit{speech} \\
    \midrule
    \textbf{True Positive} & \textbf{Sung} performance with stable pitch and melody & \textit{``We were both young when I first saw you...''} & \textit{sing} \\
    \bottomrule
    \end{tabular}
\end{table*}

\subsubsection{F1-score Calculation Protocols for SCS}
As described in Appendix \ref{app:evaluation settings for each task}, calculating the F1-score for Speech-Singing Code-Switching (SCS) requires determining whether the generated segments match the target mode (singing or speech). The determination logic for both objective (Gemini 2.5 Pro) and subjective (Human annotators) evaluations shares a common definition of singing mode but differs in the matching verification process. 

\noindent\textbf{Definition of Singing Mode.}
Both Gemini 2.5 Pro and human annotators are instructed to classify a segment as ``singing'' based purely on acoustic and melodic characteristics. The core criterion is the intent of the ``virtual speaker''. Even if the singing is imperfect (such as off-key), it is classified as the positive class as long as it exhibits a distinct melodic contour distinguishing it from speech.

\noindent\textbf{Matching Logic: Objective (F1(O)).}
For the Gemini-based evaluation, we employ an In-Context Learning (ICL) strategy, providing the model with acoustic-grounded few-shot examples to mitigate semantic bias. The process follows these steps:
\begin{enumerate}
    \item \textbf{Transcription:} Gemini transcribes the identified singing segments from the generated audio.
    \item \textbf{Segmentation:} Both the transcribed text and the ground-truth target text are segmented by commas to isolate phrases.
    \item \textbf{Fuzzy Matching:} We employ the Python \textit{thefuzz}\footnote{\url{https://www.piwheels.org/project/thefuzz/}} library to calculate the similarity between the transcribed singing segments and the target singing lyrics. A match is declared if the similarity score exceeds 70\%.
    \item \textbf{Exception for Humming:} Due to the high hallucination rate in transcription for non-lexical humming, we bypass \textit{thefuzzy} matching step for humming segments. Instead, success is determined solely by comparing the count of generated humming segments against the expected count in the target text.
\end{enumerate}

\noindent\textbf{Matching Logic: Subjective (F1(S)).}
For human evaluation, annotators directly compare the generated audio with the target text. The evaluation prioritizes timing and mode accuracy over lyrical perfection. If the model generates singing in the correct time slot but with minor lyrical deviations, it is considered a \textit{True Positive} for the mode-switching task. Such errors are penalized in WER metrics but are treated as successful mode switches. Conversely, speaking the lyrics when singing was required is treated as a mode error (\textit{False Negative}).

\subsubsection{In-Context Learning Strategy for Metric Calibration}
\label{app:icl_strategy}
Without calibration, standard multimodal LLMs exhibit a strong ``semantic bias.'' For instance, in our preliminary experiments (without ICL), we observed that baseline models like \textit{Gemini + Bark} achieved inflated scores by generating correct lyrics without singing prosody, while valid singing with ASR errors (common in \textit{Gemini + Cosy2 + LeVo}) was unfairly penalized. This discrepancy highlights that naive prompting leads the evaluator to rely on transcribed semantics rather than acoustic realization. To correct this and ensure F1(O) accurately reflects prosodic mode-switching, We implemented a rigorous In-Context Learning  strategy comprising two key components: task-specific system instructions and contrastive acoustic demonstrations.

\begin{figure*}[t] 
    \centering
    \includegraphics[scale=0.5]{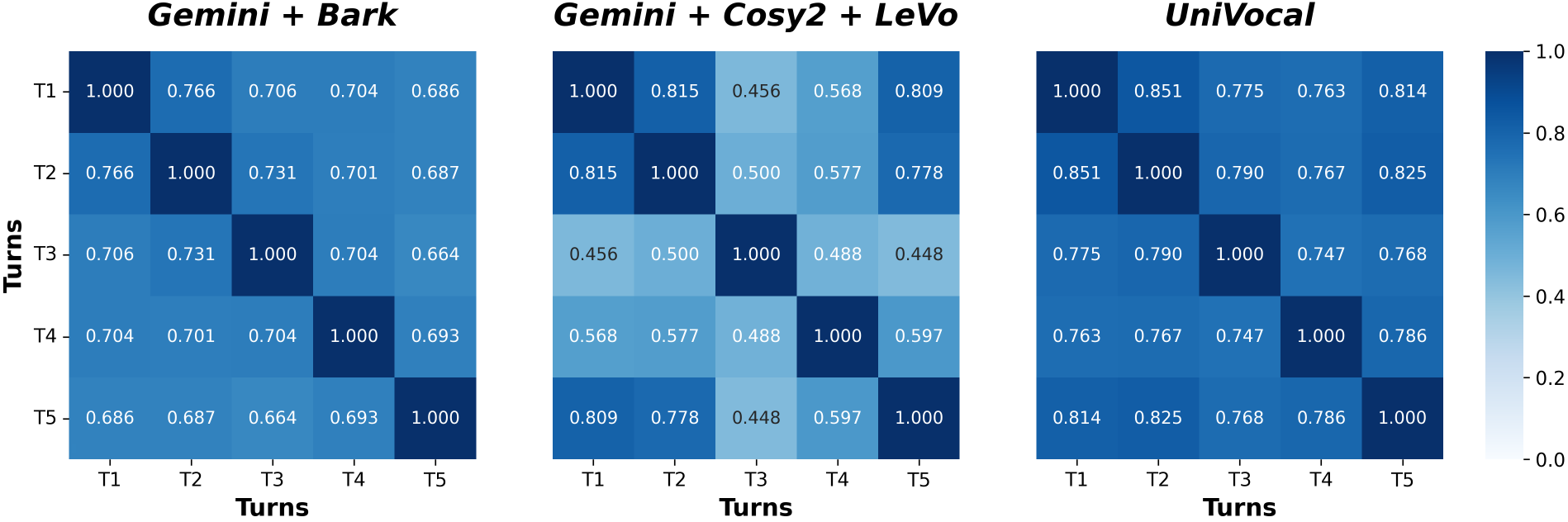}

    \caption{ Intra-sample speaker consistency. Pairwise similarity heatmap between five temporal segments, averaged across all generated samples from each system. Darker colors indicate higher speaker stability.}
    \label{fig:app_sim_of_turns}

\end{figure*}

\begin{table*}[t]
 \centering
 \small
 \caption{Results on the textual empathy task. Metrics suffixed with (O) and (S) denote evaluations by Gemini 2.5 Pro and human annotators, respectively. The \textbf{bold} and \underline{underlined} numbers indicate the optimal and sub-optimal results, respectively.}
 \label{table:app_empathy}
 \begin{tabular}{l c c c c c} 
  \toprule
  \multirow{2}{*}{\textbf{Model}} & \multicolumn{5}{c}{\textbf{Textual Empathy Test Set}} \\
  \cmidrule(lr){2-6}
   & E-MOS(O)$\uparrow$ & P-MOS(O)$\uparrow$ & E-MOS(S)$\uparrow$ & P-MOS(S)$\uparrow$ & WER$\downarrow$ \\  
  \midrule
   CosyVoice 2 & 1.66 & 1.93 & 1.78 & 1.74 & 0.53 \\
   {ElevenLabs' multilingual-v2} & \textbf{2.26} & \textbf{2.63} & \textbf{2.30} & \textbf{2.47} & \textbf{0.24} \\
  \midrule
   \textit{UniVocal} & \underline{2.11} & \underline{2.25} & \underline{2.26} & \underline{2.22} & \underline{0.32} \\ 
  \bottomrule
 \end{tabular}
\end{table*}

\noindent\textbf{Task-Specific System Instructions.}
We constructed a structured system prompt that explicitly defines the segmentation rules and labeling conventions. Crucially, the prompt includes a \textit{Negative Constraint} to counteract semantic bias:
\begin{quote}
    \textit{``Focus exclusively on acoustic features... You must distinguish `speech' from `singing' based ONLY on prosody, pitch modulation, rhythm, and melody, NOT on the text content. Ignore semantic cues: Do not classify a segment as `singing' just because the lyrics look like a song.''}
\end{quote}
The model is instructed to output a structured log containing timestamps and transcriptions, labeled strictly as \textit{speech}, \textit{sing}, or \textit{hum}.

\noindent\textbf{Contrastive Acoustic Demonstrations (Few-Shot).}
To further ground the model's understanding in acoustics, we provided two carefully curated 1-shot examples (demonstrations) representing a ``Hard Negative'' and a ``True Positive'' case. As illustrated in Table \ref{tab:icl_examples}, these examples use identical or semantically similar lyrical content but differ significantly in vocal mode:

\begin{itemize}
    \item \textbf{Hard Negative Example (Speech):} We selected an audio sample where a speaker recites the lyrics of a famous song (\textit{Love Story}) in a spoken prosody. Despite the text being clearly lyrical, the ground-truth label provided to Gemini is \textit{speech}. This forces the model to override its textual prior and attend to the flat pitch and lack of musical tempo.
    \item \textbf{True Positive Example (Singing):} We selected a genuine singing sample of the same song. The label provided is \textit{sing}, reinforcing the association between the label and acoustic features like sustained pitch and melodic intervals.
\end{itemize}

By processing these contrastive examples in the context window before inference, Gemini 2.5 Pro learns to decouple semantic content from vocal mode, significantly improving its reliability as an automated evaluator for the SCS task.

\subsubsection{Automated Metric Calibration Analysis}
\label{app:f1_calibration}
To validate the reliability of Gemini 2.5 Pro as an automated evaluator, we conducted a correlation analysis against human ratings on 243 generated audio samples. We observed a statistically positive correlation at the sample level (Pearson $r=0.343$, Spearman $\rho=0.346$, $p<0.05$). 

While the magnitude of these coefficients indicates moderate correlation, it is inherently limited by the discrete nature of F1 scores on short samples. Since sample-level F1 scores often result in binary outcomes (0.0 or 1.0) due to the ability of baselines, the lack of continuous variance dampens linear correlation metrics. However, when aggregating these scores to the system level, the noise averages out. As detailed in the main text, the automated metric exhibits perfect rank consistency with human evaluation across all \textbf{SCSBench} subsets, confirming its validity as a proxy for distinguishing relative system performance.

\begin{table*}[htbp]
\centering
\small
\caption{Objective performance across different cent token resolutions.}
\begin{tabular}{lccccc}
\toprule
& \multicolumn{3}{c}{\textbf{Textual Empathy Test Set}} & \multicolumn{2}{c}{\textbf{Fullsong Test Set}} \\
\cmidrule(lr){2-4} \cmidrule(lr){5-6}
\textbf{Resolution} & \textbf{E-MOS(O)$\uparrow$} & \textbf{P-MOS(O)$\uparrow$} & \textbf{WER$\downarrow$} & \textbf{AES$\uparrow$} & \textbf{WER$\downarrow$} \\
\midrule
12 bins & 1.57 & 1.63 & 0.46 & \textbf{3.61} & 58.87\% \\
480 bins & 1.82 & 1.97 & 0.51 & 3.42 & \textbf{49.72\%} \\
1200 bins & \textbf{1.85} & \textbf{2.06} & \textbf{0.42} & 3.45 & 56.13\% \\
\bottomrule
\end{tabular}
\label{tab:resolution_ablation}
\end{table*}

\begin{table}[htbp]
\centering
\small
\caption{Correlation between predicted and GT cent tokens.}
\begin{tabular}{lcc}
\toprule
\textbf{Dataset} & \textbf{SRCC} & \textbf{LCC} \\
\midrule
Textual Empathy Test Set & 0.633 & 0.604 \\
Fullsong Test Set & 0.679 & 0.628 \\
\bottomrule
\end{tabular}
\label{table:pitch plan}

\end{table}

\subsubsection{MOS Scoring Criteria for Other Tasks}
The Mean Opinion Score (MOS) evaluations utilize a 3-point scale (1.0 to 3.0, in increments of 0.25). While E-MOS and P-MOS are applied identically in both Gemini-based objective and human subjective evaluations, N-MOS and M-MOS are assessed exclusively by human annotators.

\noindent\textbf{E-MOS} 
\begin{itemize}
    \item \textbf{1.0 (Poor):} The delivery is flat, monotonous, or emotionally neutral, showing no connection to the text's emotional content.
    \item \textbf{2.0 (Acceptable):} An attempt at emotion is audible, but it is either weak, inconsistent, or does not fully match the nuance of the text.
    \item \textbf{3.0 (Excellent):} The emotional delivery is clear, authentic, and perfectly aligns with the emotion and intensity implied by the text. The performance is convincing.
\end{itemize}

\noindent\textbf{P-MOS} 
\begin{itemize}
    \item \textbf{1.0 (Poor):} Delivery is unclear, monotonous, or robotic. Words may be slurred. Pausing, rhythm, and stress are unnatural and hinder comprehension.
    \item \textbf{2.0 (Acceptable):} Delivery is generally clear but may have minor flaws. Prosody is understandable but may lack naturalness.
    \item \textbf{3.0 (Excellent):} Delivery is exceptionally clear, fluid, and natural. Intonation, rhythm, and stress effectively enhance the meaning. Every word is crisply enunciated.
\end{itemize}

\noindent\textbf{N-MOS} 

\begin{itemize}
    \item \textbf{1.0 (Poor):} The voice sounds distinctly synthesized, metallic, or robotic. Contains significant acoustic artifacts, glitches, or unnatural phase issues.
    \item \textbf{2.0 (Acceptable):} The voice sounds generally human-like but exhibits occasional artifacts or slight unnaturalness in timbre that reveals its synthetic nature.
    \item \textbf{3.0 (Excellent):} The voice is indistinguishable from a real human recording. The timbre is rich, warm, and free of any audible processing artifacts.
\end{itemize}

\noindent\textbf{M-MOS} 
\begin{itemize}
    \item \textbf{1.0 (Poor):} Significant off-key notes, unstable pitch, or erratic rhythm. The singing is unpleasant to listen to or musically incoherent.
    \item \textbf{2.0 (Acceptable):} Pitch and rhythm are mostly correct, but the performance lacks expressiveness or "soul." It is technically adequate but musically flat or generic.
    \item \textbf{3.0 (Excellent):} The performance is musically engaging with stable pitch and precise rhythm. It demonstrates musical nuance that enhances aesthetic appeal.
\end{itemize}

\section{Additional Experimental Details and Results}

\begin{table*}[htbp]
\centering
\small
\caption{Mode-switching F1 scores on real-world SCS scenarios.}
\begin{tabular}{lcc}
\toprule
\textbf{Model} & \textbf{Real SCS (F1)} & \textbf{Enhanced Real SCS (F1)} \\
\midrule
Gemini + Cosy2 + LeVo & 0.452 & 0.691 \\
UniVocal & 0.201 & 0.730 \\
\bottomrule
\end{tabular}
\label{tab:real_world_scs}
\end{table*}

\subsection{Extended Analysis on Intra-Sample Speaker Consistency}
To further investigate the stability of speaker identity during mode transitions, we extend the analysis presented in Section \ref{sec:main result} by including the \textit{Gemini + Bark} baseline in our intra-sample speaker consistency evaluation. We utilize the same metric methodology: partitioning generated samples into five temporal segments and computing the average pairwise speaker similarity between all segments to visualize identity drift. The comparative results are illustrated in Figure \ref{fig:app_sim_of_turns}.

A critical divergence appears when contrasting intra-sample consistency with the global metrics. Although \textit{Gemini + Cosy2 + LeVo} achieves the highest global speaker similarity (SIM in Table \ref{table:mix_results_quality}) , its intra-sample speaker stability is compromised by the use of distinct models for speech and singing, resulting in noticeable timbre mismatches between segments. Conversely, \textit{UniVocal} and \textit{Gemini + Bark} exhibits superior internal coherence due to their unified architecture. By effectively balancing precise switching timing (F1-scores in Table \ref{table:mix_results_f1}), competitive global similarity, and robust intra-sample consistency, \textit{UniVocal} stands out as the optimal framework for high-quality speech-singing code-switching synthesis.

\subsection{Additional Results on Textual Empathy}
Table \ref{table:app_empathy} presents the comprehensive evaluation results on the textual empathy test set, comparing objective scores from Gemini 2.5 Pro (suffixed with (O)) against the subjective human ratings (suffixed with (S)) reported in the main text. Notably, while slight deviations exist in absolute values, the system-level rankings remain consistent across both evaluators: \textit{ElevenLabs} $>$ \textit{UniVocal} $>$ \textit{CosyVoice 2}. This strong alignment with human preference confirms that Gemini 2.5 Pro captures the nuances of emotional expression and prosody effectively, validating its potential as a reliable automated evaluator for expressive TTS.

\subsection{Ablation on Cent Token Resolution}
\label{app:CoT resolution}
We conducted a supplementary ablation study to evaluate the impact of different cent token resolutions on generation quality. We trained model variants with 12 bins, 480 bins, and the default 1200 bins. Due to computational constraints, each variant was trained with the Text-to-Vocal LLM for 3 epochs and Flow Matching for 1 epoch.

The objective results in Table \ref{tab:resolution_ablation} indicate that a higher token resolution yields markedly better performance in expressive speech while maintaining competitive results in singing tasks. The 12-bin setting, while sufficient for basic melody, is too coarse to capture the micro-prosody required for empathetic speech, leading to significantly lower E-MOS and P-MOS scores. Thus, our choice of 1200 bins provides the optimal granularity for the unique demands of \textit{UniVocal}.

\subsection{The Prosodic Planning Mechanism}
\label{app:pitch plan}
To verify whether the refined cent token truly acts as a structural planner rather than merely providing supplemental acoustic details, we conducted experiments to measure the alignment between model predictions and acoustic reality. We extracted the generated refined cent tokens during inference and calculated their correlation against the Ground Truth (GT) cent tokens—extracted directly from the final synthesized audios—using SRCC and LCC.

As shown in Table \ref{table:pitch plan}, the positive correlations demonstrate that the generated cent tokens effectively outline the primary pitch contour, upon which subsequent semantic tokens supplement finer details. This confirms that \textit{UniVocal} does indeed draft a structural pitch framework before generating linguistic content, fully realizing a prosodic "planning" mechanism. We have also updated our anonymous demo page with side-by-side visual plots comparing predicted vs. GT cent tokens.

\subsection{Performance on Real-World SCS Scenarios}
\label{app:real SCS}
To validate \textit{UniVocal}'s generalization capabilities beyond synthetic texts, we collected approximately 30 minutes of real-world human SCS recordings from the internet. Since our training SCS data predominantly consists of speech segments, we trimmed the singing portions of the real-world data to maintain a similar distribution, forming the Real SCS test set. Additionally, we created an Enhanced Real SCS test set by manually inserting one explicit semantic cue into the text context of each sample.

As shown in Table \ref{tab:real_world_scs}, UniVocal initially faces challenges with the domain gap inherent in real-world SCS scenery (F1 of 0.201). However, its performance surges to 0.730 on the Enhanced set. This demonstrates that with the addition of minor explicit semantic triggers acting as anchors, UniVocal generalizes remarkably well to real-world scenarios, achieving a mode-switching accuracy close to its in-domain performance.

\end{document}